\documentstyle[proceedings,numreferences]{crckapb}
\input psfig
\pssilent
\begin{opening}
\title{Numerical Relativity:}
\subtitle{Towards Simulations of 3D Black Hole Coalescence}
\author{Edward Seidel}
\institute{Max-Planck-Institut-f\"ur-Gravitationsphysik\\
Schlaatzweg 1, 14473 Potsdam, Germany
}
\end{opening}
\begin{document}
\begin{abstract}
I review recent developments in numerical relativity, focussing on 
progress made in 3D black hole evolution.  Progress in development of 
black hole initial data, apparent horizon boundary conditions, 
adaptive mesh refinement, and characteristic evolution is highlighted, 
as well as full 3D simulations of colliding and distorted black holes.  
For true 3D distorted holes, with Cauchy evolution techniques, it is 
now possible to extract highly accurate, nonaxisymmetric waveforms 
from fully nonlinear simulations, which are verified by comparison to 
pertubration theory, and with characteristic techniques extremely long 
term evolutions of 3D black holes are now possible.  I also discuss 
a new code designed for 3D numerical relativity, called Cactus, that 
will be made public.
\end{abstract}
\vspace{-10.5cm} 
\begin{flushright}
\baselineskip=15pt

\end{flushright}
\vspace{9cm}

\section{Introduction}

Numerical Relativity is having broad impact across many areas of 
relativity, astrophysics, and cosmology.  Because of the pervasiveness 
of numerical techniques in relativity, it is simply impossible to 
survey the entire field in a plenary talk.  Therefore, I will focus on 
a single area that cuts across many of these fields, and one which has 
galvanized the numerical relativity community: black holes (BH's).  
This particular research illustrates many of the issues facing numerical 
relativists very well.  Just to preview my overview of this subject, 
here is how I see the current status:

{\em The Need.} We need full 3D numerical relativity for gravitational 
wave astronomy.  The imminent arrival of data from of the long awaited 
gravitational wave interferometers (see, e.g., Ref.~\cite{Flanagan97b} 
and references therein) has provided a sense of urgency in producing 
realistic simulations of strong sources of gravitational waves, 
possible only through the full machinery of numerical relativity.  As 
has been emphasized by Flanagan and Hughes, one of the best candidates 
for early detection by the laser interferometer network is 
increasingly considered to be BH 
mergers\cite{Flanagan97b,Flanagan97a}.  However, the signals are 
likely to be weak enough by the time they reach the detectors that 
reliable detection may be difficult without prior knowledge of the 
merger waveform.  Flanagan's talk in this volume reviews these issues 
in detail.  These are among the reasons that the NSF-funded Binary 
Black Hole Grand Challenge Alliance has focused the efforts of 
numerous US and international groups on developing codes for solving 
the problem of 3D coalescing BH's.  

{\em The Problems.} There are many technical problems that must be 
solved before we can perform realistic simulations of BH merger events 
that will be useful for gravitational wave astronomy.  I will provide 
a status report on the following issues: ({\em a}) The initial value 
problem.  One must have initial data representing two astrophysically 
relevant BH's orbiting each other in order to begin a simulation.  
({\em b}) Boundary conditions.  In any numerical code (with a finite 
boundary), boundary conditions are essential, and this is particularly 
true of the BH problem.  Both the inner boundary, (say, inside 
the event horizon), and the outer boundary are problematic.  ({\em c}) 
Adaptive mesh refinement.  The computations of 3D relativity are so 
demanding that even on the world's largest computers, one will have to 
resort to clever techniques to resolve numerically only those 
spacetime regions that demand it, or else the calculations will be 
intractable.  Adaptive mesh refinement is being developed to refine 
the calculations only where it is needed.

{\em The Goal: Waveforms.} There are many reasons to pursue numerical 
relativity, even within the area of BH collisions (e.g. 
theoretical studies of the event horizons of dynamic BH's can 
now be made through numerical 
relativity\cite{Anninos94f,Matzner95a,Shapiro95a}).  However, for 
gravitational wave astronomy, a most important goal of numerical 
relativity is the calculation of waveforms expected from the inspiral 
and merger.  We will see that accurate waveforms from nonaxisymmetric 
BH simulations are already possible, even if they carry only a 
tiny fraction of the ADM mass in energy.

{\em The Codes: Focusing Large Scale Efforts.} In order to make real 
progress in 3D numerical relativity, one needs many skills.  A wide 
range of difficult problems face us, ranging from mathematical 
formulations of the equations to advanced computational science 
techniques on parallel computers.  Yet in the end a simulation must be 
performed by a single evolution code.  For this reason, the efforts of 
many groups around the world have been focussed on the development of 
a small number of evolution codes.  I will focus on one such 3D code, 
called Cactus, that is being used in many different projects, and will 
be made available to the community soon.

{\em The Future: BH's, Neutron Stars, The Universe.} With so much 
activity on the rather narrow subject of BH's to report on, 
there is unfortunately no room to discuss many other exciting areas in 
numerical relativity, such as critical phenomena, neutron star 
evolutions, and cosmology.  But in summary, progress in this field is 
excellent, and we can look forward to many discoveries through 
numerical approaches to relativity in the future.

\section{Initial Value Problem}
In this section I review briefly the status of solving the initial 
value problem for BH's.  As with any initial data for Cauchy 
evolution in numerical relativity, the basic idea is to find relevant 
solutions to the Hamiltonian and momentum constraints that contain 
BH's, and evolve them.  As we will see in this section, the key 
difficulty lies in the word ``relevant''; we now have at our disposal 
techniques to generate far more complicated datasets than we have the 
capability to actually evolve numerically.  

I will not have space to review the formalism for developing initial 
data for numerical relativity.  The standard article for this is still 
York's classic\cite{York79}.  (For relevant BH overviews, see 
also\cite{Seidel96a,Seidel96b,Seidel97a,Seidel97b,Seidel98a}.)  For 
notational purposes, the 3--metric is generally written as $ds^{2} = 
\psi^{4} \hat{ds}^{2}$ where $\hat{ds}^{2}$ is a known metric (often 
chosen to be the flat metric), and $ds^{2}$ is the unknown metric for 
which we are solving.  Then the hamiltonian constraint is written as 
an elliptic equation for the unknown conformal factor $\psi$, which 
can be solved, given a solution for the extrinsic curvature $K_{ij}$ 
to the momentum constraints (e.g. time symmetric data, or $K_{ij}=0$).  
Once these data are given, they must be evolved, given a choice of 
lapse and shift.

\subsection{Schwarzschild and Distorted Schwarzschild}
The BH dataset most familiar to all relativists is the 
Schwarzschild solution.  Although this spherical BH solution 
is now more than 80 years old, it is still an important solution to 
the constraints that is being used to test numerical relativity codes.  
When written in the notation of 3D numerical relativity, the 3--metric 
becomes
\begin{equation}
	ds^{2} = \psi^{4} \hat{ds}^{2}\\
	= \psi^{4} (dx^{2} + dy^{2} + dz^{2})\\
		=(1 + M/2r)^{4}(dx^{2} + dy^{2} + dz^{2})
\end{equation}
where $r$ is the standard isotropic radius.  This solution is still 
very relevant today, as any bound BH system without angular 
momentum (e.g., two BH's colliding head on) must settle towards 
this solution at late times.  With the standard Schwarzschild lapse 
this metric is the solution for all time, but with a dynamic slicing 
the 3--metric will evolve.

Now, imagine two BH's colliding violently: merging at nearly the speed 
of light, their horizons combine to form a single, highly distorted 
BH. This BH must then settle down to its final equilibrium state.  The 
Schwarzschild dataset was generalized to include such highly 
distorted, dynamic BH's by numerous researchers, beginning in the 
1980's by Bernstein, Hobill, and Smarr.  These datasets have been 
evolved in axisymmetry for a decade, and are now finding their 
way into full 3D simulations.  They are very useful, since they allow 
one to explore the dynamics of distorted BH's, such as those that will 
be formed during black hole collisions, without having to first evolve 
the inspiral.  One simply starts with a distorted ``Schwarzschild'' 
(i.e., non-rotating) or ``Kerr'' (i.e., rotating) BH as initial data.

These datasets correspond to a gravitational wave of the form 
originally considered by Brill\cite{Brill59} superimposed on 
Schwarzschild.  The flat conformal 3--metric $\hat 
ds^{2}$ is replaced by the ``Brill'' form with adjustable 
gravitational wave parameters.  Such data sets mimic the state of two 
BH's colliding, and form a useful model for studying the late stages 
of BH coalescence.

The 3--metric is $d\ell^2 = \tilde{\psi}^4 \left( e^{2q} \left( 
d\eta^2 + d\theta^2 \right) + \sin^2\theta d\phi^2 \right), $where 
$\eta$ is a radial coordinate related to the Cartesian coordinates by 
$\sqrt{x^{2}+y^{2}+z^{2}} = e^{\eta}$.  For details, please 
see\cite{Camarda97b}.  Given a choice for the ``Brill wave'' function 
$q$, the Hamiltonian constraint leads to an elliptic equation for the 
conformal factor $\tilde{\psi}$.  The function $q$ represents the 
gravitational wave surrounding the BH, and can be chosen freely to 
give a variety of distortion amplitudes and shapes (with some 
restrictions.)  If the Brill wave amplitude vanishes, the undistorted 
Schwarzschild solution results, and for small amplitudes, the data 
corresponds to a perturbed BH. These data sets can also include 
angular momentum \cite{Brandt97a,Brandt97c}, in which case the momentum 
constraints must also be solved.  The rotating versions of these 
datasets build on the original rotating datasets of Bowen and 
York~\cite{Bowen80}, which are contained as subsets of these more 
general datasets.  Together, these datasets form a rich testing ground 
for BH evolution codes designed to treat the coalescence 
problem, as well as a laboratory for studying the dynamics of 
distorted BH's.  We will see results of evolutions of such 
BH data below.

\subsection{Multiple BH Data}
The datasets described above all have an Einstein-Rosen bridge 
construction: a simple wormhole connecting two identical 
asymptotically flat sheets.  Such constructions were generalized over 
30 years by Misner\cite{Misner60}, Brill, Lindquist\cite{Brill63} and 
others to include two wormholes, leading to what we now know as two 
BH initial data.  The Misner solution corresponds to two 
axisymmetric, equal mass BH's, initially at rest (time 
symmetric initial data: $K_{ij}=0$).  This is a single parameter 
family of initial data with an adjustable distance between the 
wormholes. 

This family of initial data has become something of a classic in 
numerical relativity: the first attempt to evolve it numerically was 
by Hahn and Lindquist in 1963\cite{Hahn64}, even before the modern 
notions of BH's or the ADM formalism had been fully developed.  
In the 1970's DeWitt gave the problem to his student Larry Smarr, and 
along with \v{C}ade\v{z} and Eppley more modern numerical methods and slicing 
conditions were applied to the problem, this time with some 
success\cite{Smarr77}.  Again in the 1990's, the same initial data 
were evolved again, this time with more powerful computers and 
numerical techniques, and at last reliable waveforms could be 
determined.  This modern work also helped spark a renaissance of 
perturbative approaches to the problem, as outlined by Pullin in his 
plenary lecture.  In sections below I will review recent numerical 
results in both axisymmetry and 3D. But the bottom line is that even 
these most simple possible BH collisions are still very 
challenging problems that continue to stress the most advanced 
numerical codes and computers we have!

However, we are ultimately interested in solving the more general 3D 
BH coalescence problem, with different masses, and with spin and 
orbital angular momentum.  Techniques to create such initial datasets 
were developed by York and colleagues, especially Greg Cook.  
Generalizing the original ideas of Misner to create multiple wormhole 
datasets with two identical asymptotically flat sheets (i.e., there 
exists an isometry operator through the ``throats'' of the wormholes, 
mapping the top sheet to an identical one below), one can now generate 
full 3D datasets by solving both the momentum and Hamiltonian 
constraints\cite{Cook93}.  A series of such initial datasets has been 
analyzed by Cook\cite{Cook94}.  Generally the numerical solution is 
found only on one sheet, with the isometry operator providing boundary 
conditions on the throat.  Mathematically straightforward, this can be 
painful to implement in 3D cartesian coordinates!  An important 
variation on these techniques is the Brandt-Br{\"u}gmann 
construction\cite{Brandt97b}, which was only developed last year and 
evolved for the first time.  Rather than an isometry surface, through 
which one universe is mapped to an identical one ``below'', it has a 
singularity inside each hole that is built-in analytically.  The 
numerical solution, for the nonsingular part, is then regular on the 
entire domain, which is very convenient to solve for in 3D cartesian 
coordinates.

The bottom line is that we have more initial sets than we can evolve 
right now!  Full 3D data sets are ready, and waiting for us!  However, 
the problems of evolution are far more difficult, as I will outline 
below.  But even about the initial data, there is still a major 
caveat: although we can now generate very accurate binary BH initial 
data, with arbitrary spin and momenta, we really do not understand 
their connection to astrophysics well.  The initial data will contain 
some gravitational wave content over which we have little control.  
Furthermore, how to match a given initial dataset to a particular 
inspiral scenario is unknown at present.  So there is still much to be 
done even at the level of providing astrophysically relevant initial 
data.

\section{The trouble with black holes}

As I have described at length, we have many BH datasets at our 
disposal for evolution.  But they all have in common one problem: 
singularities lurk within them, which must be handled numerically.  
Developing suitable techniques for doing so is one of the major 
research priorities of the community at present.  If one attempts to 
evolve directly into the singularity, infinite curvature will be 
encountered, causing any numerical code to break down.

\begin{figure}
\centerline{ \psfig{figure=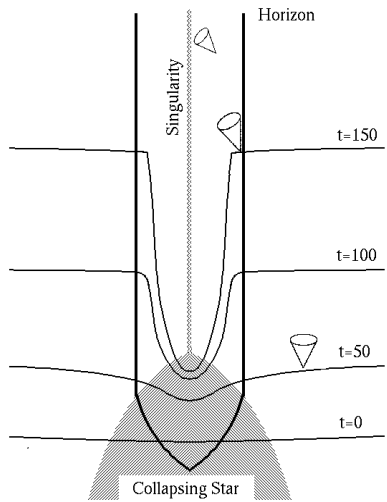,height=70mm}} \caption{A spacetime 
diagram showing the formation of a BH, and time slices 
traditionally used to foliate the spacetime in traditional numerical 
relativity with singularity avoiding time slices.  As the evolution 
proceeds, pathologically warped hypersurfaces develop, leading to 
unresolvable gradients that cause numerical codes to crash.}
\label{singularity}
\end{figure}

Traditionally, the singularity region is avoided by the use of 
``singularity avoiding'' time slices, that wrap up around the 
singularity.  Consider the evolution shown in Fig.~\ref{singularity}.  
A star is collapsing, a singularity is forming, and time slices are 
shown which avoid the interior while still covering a large fraction 
of the spacetime where waves will be seen by a distant observer.  
However, these slicing conditions by themselves do not solve the 
problem; they merely serve to delay the onset of instabilities.  As 
shown in Fig.~\ref{singularity}, in the vicinity of the singularity 
these slicings inevitably contain a region of abrupt change near the 
horizon, and a region in which the constant time slices dip back deep 
into the past in some sense.  This behavior typically manifests itself 
in the form of sharply peaked profiles in the spatial metric 
functions~\cite{Smarr78b}, ``grid stretching''~\cite{Shapiro86} or large 
coordinate shift~\cite{Bernstein89} on the BH throat, etc.  
Numerical simulations will eventually crash due to these pathological 
properties of the slicing.

\subsection{Apparent Horizon Boundary Conditions (AHBC)}
Cosmic censorship suggests that in physical situations, singularities 
are hidden inside BH horizons.  Because the region of 
spacetime inside the horizon is causally disconnected from the region 
of interest outside the horizon, one is tempted numerically to cut 
away the interior region containing the singularity, and evolve only 
the singularity-free region outside, as originally suggested by 
Unruh\cite{Unruh84}.  This has the consequence that there will be a 
region inside the horizon that simply has no numerical data.  To an 
outside observer no information will be lost since the regions cut 
away are unobservable.  Because the time slices will not need 
such sharp bends to the past, this procedure will drastically reduce 
the dynamic range, making it easier to maintain accuracy and 
stability.  Since the singularity is removed from the numerical 
spacetime, there is in principle no physical reason why BH 
codes cannot be made to run indefinitely without crashing.

We spoke innocently about the BH horizon, but did not 
distinguish between the {\em apparent} and {\em event} horizon.  These 
are very different concepts!  While the event horizon, which is 
roughly a null surface that never reaches $\cal{I}$ and never hits the 
singularity, may hide singularities from the outside world in many 
situations, there is no guarantee that the apparent horizon, which is 
the (outermost) surface that has instantaneously zero expansion 
everywhere, even exists on a given slice!  While methods for finding 
event horizons in numerical spacetimes are now known, and have been 
used to determine much interesting physics, they can only be found 
after examining the {\em history} of an evolution that has been 
already been carried out to sufficiently late 
times\cite{Anninos94f,Libson94a}.  Hence they are useless in 
providing boundaries as one integrates {\em forward} in time.  On the 
other hand the apparent horizon, if it exists, can be found on any 
given slice by searching for closed 2--surfaces with zero expansion.  
Although one should worry that in a generic BH collision, one 
may evolve into situations where no apparent horizon actually exists, 
let us cross that bridge if we come to it!  

Given these considerations, there are two basic ideas behind the 
implementation of the apparent horizon boundary condition:

{\em (a)} It is important to use a finite differencing scheme which 
respects the causal structure of the spacetime.  Since the horizon is 
a one-way membrane, quantities on the horizon can be affected only by 
quantities outside but not inside the horizon: all quantities on the 
horizon can in principle be updated solely in terms of known 
quantities residing on or outside the horizon.  There are various 
technical details and variations on this idea, which is called 
``Causal Differencing''\cite{Seidel92a} or ``Causal 
Reconnection''\cite{Alcubierre94a}, but here I focus primarily on the 
basic ideas and results obtained to date.

{\em (b)} A shift is used to control the motion of the horizon, and 
the behavior of the metric functions outside the BH.

An additional advantage to using causal differencing is that it allows 
one to follow the information flow to create grid points with proper 
data on them, as needed inside the horizon, even if they did not exist 
previously.  (Remember above that we have cut away a region inside the 
horizon, so in fact we have no data there.)  This process has been 
termed ``educating grid points before birth'' by Wai-Mo Suen.  This 
will be an important education if one wants to let a BH move 
across the computational grid.  If a BH is moving physically, 
it is also desirable for it to move through coordinate space.  
Otherwise, all physical movement will be determined by metric function 
evolution.  For a single BH moving in a straight line, this 
may be reasonable, but for spiraling coalescence this will lead to 
hopelessly contorted grids.  The immediate consequence of this is that 
as a BH moves across the grid, regions in the wake of the 
hole, now in its exterior, must have previously been inside it where 
no data exist!  But with AHBC and causal differencing this need not be 
a problem.

Does the AHBC idea work?  Preliminary indications are very promising.  
In spherical symmetry (1D), numerous studies show that one can 
successfully locate horizons, cut away the interior, and evolve for 
essentially unlimited times ($t \propto 10^{3-4}M$).  The growth of 
metric functions can be completely controlled, errors are reduced to a 
very low level, and the results can be obtained with a large variety 
of shift and slicing conditions, and with matter falling in the BH to 
allow for true dynamics even in spherical 
symmetry\cite{Seidel92a,Anninos94e,Scheel94,Marsa96}.

In 3D, the basic ideas are similar but the implementation is much more 
difficult.  The first successful test of these ideas to a 
Schwarzschild BH in 3D used horizon excision and a shift provided from 
similar simulations carried out with a 1D code\cite{Anninos94c}.  The 
errors were found to be greatly reduced when compared even to the 1D 
evolution with singularity avoiding slicings.  (Another 3D 
implementation of the basic technique was provided by 
Br{\"u}gmann~\cite{Bruegmann96}.)

This was a proof of principle, but more general treatments are 
following.  In collaboration with the NCSA/WashU group, Daues extended 
this work to a full range of shift conditions~\cite{Daues96a}, 
including the full 3D minimal distortion shift~\cite{York79}.  He also 
applied these techniques to dynamic BH's, including Misner data (where 
the holes are close enough together to be a single distorted 
Schwarzschild hole initially), and collapse of a 3D boson star to form 
a BH, at which point the horizon is detected, the region interior to 
the horizon excised, and the evolution continued with AHBC. The focus 
of this work has been on developing general gauge conditions for 
single BH's without movement through a grid.  Under these conditions, 
BH's have been accurately evolved well beyond $t=100M$.

\begin{figure}
\centerline{ \psfig{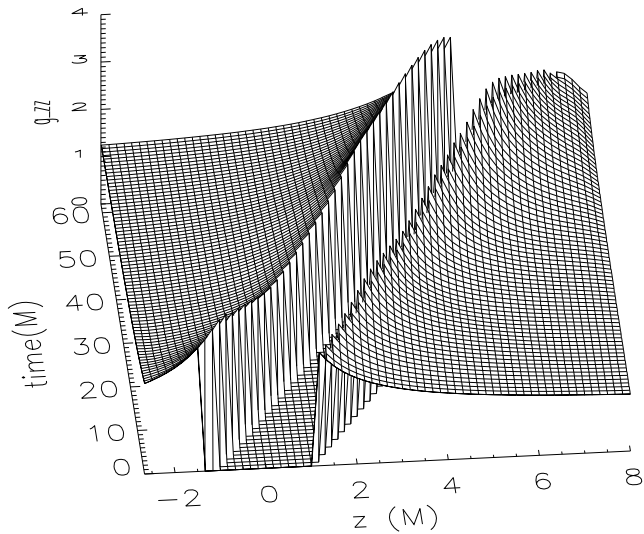}} 
\caption{Metric component $g_{zz}$ along the $z$-axis is shown as a function of 
time for a boosted Schwarzschild BH evolved with AHBC. The flat region 
that moves diagonally to the right represents the excised region 
(inside the black hole).  Note that points at the trailing edge (left 
side) are smoothly updated as the hole moves towards positive $z$.}
\label{moving}
\end{figure}

Taking the approach in a different direction, work of the Grand 
Challenge Alliance has been focussed on development of 3D AHBC 
techniques for {\em boosted} Schwarzschild BH's\cite{Cook97a}.  In 
this work, analytic gauge conditions are provided, which are chosen to 
make the evolution static, although the numerical evolution is allowed 
to proceed freely.  The boosted hole allows the first test of Suen's 
``education of grid points before birth'' as they emerge in the BH 
wake.  Using causal differencing, this effort has successfully moved 
the BH several diameters across the grid, and accurate evolutions have 
now been carried out for $t \approx 500M$.  In Fig.~\ref{moving}, 
recent results from such experiments are shown.

These new results are significant achievements, and show that the 
basic techniques outlined above are not only sound, but are also 
practically realizable in a 3D numerical code.  However, there is 
still a significant amount of work to be done!  The techniques have 
yet to be applied carefully to distorted BH's, with tests of 
the waveforms emitted (see below), they have not be applied to 
rotating BH's of any kind, they have not been applied to 
colliding BH's with horizon topology change, and moving black 
holes have yet to be evolved in AHBC with a nonanalytic gauge choice.  
There are still clearly many steps to be taken before the techniques 
will be successfully applied to the general BH merger problem.

\section{Characteristic Evolution of 3D BH's}
Another very recent approach to 3D BH evolution that completely avoids 
the problems of grid stretching is characteristic evolution.  The 
Pittsburgh group, in collaboration with the Grand Challenge Alliance, 
has developed the first full 3D characteristic code evolving nonlinear 
Einstein equations.  This technique was originally envisioned as an 
approach to the problem of computing the spacetime in the far zone of 
the BH, where it would be matched to an interior Cauchy evolution code 
(Cauchy-Characteristic matching).  In such an application, the 
characteristic portion of the spacetime would be foliated by outgoing 
null surfaces so that essentially outgoing radiation would be carried 
away to $\cal{I}$, but in this case it has been applied to the problem 
of evolving the BH's themselves\cite{Gomez97a,Gomez98a}.  The code 
uses the Bondi-Sachs form of the metric, and in the BH application 
evolves a region of spacetime from a region about $10M$ outside the 
horizon to the horizon itself, foliated by {\em ingoing} 
characteristic slices.

Using this technique, the characteristic code has successfully evolved 
3D BH's for essentially unlimited times ($t \approx 60,000M$).  The 
results are even more impressive when one considers the fact that not 
only Schwarzschild BH's were evolved, but also distorted and rotating 
BH's.  To my knowledge these are the first rotating BH's to be evolved 
in 3D. The distorted BH's consist of radiation imposed on the initial 
ingoing null surface, which then propagates in, hits the BH, and for the 
most part enter the horizon.

However, it seems likely that this method by itself will encounter 
difficulties for evolution of very highly distorted or colliding black 
holes, where focusing of ingoing light rays may create caustics, 
leading to a breakdown of the foliation.  Also, ironically, the method 
is presently most successful when a BH {\em is} present, creating an 
$S^{2} \times R$ topology; dealing with the so-called $r=0$ problem is 
difficult for any formulation of the Einstein equations, and is 
avoided by using cartesian grids in the standard 3+1 formulations, but 
the characteristic method does not use cartesian grids, and would 
therefore have to face this problem in the absence of a BH (e.g., for 
the coalescence of neutron stars).  Nonetheless, the possibility of 
very long time evolutions demonstrated with the characteristic 
evolution scheme is an exceptional achievement that seems likely to 
provide an alternate and superior approach for an interesting class of 
3D BH spacetimes.  It also provides strong evidence that 
characteristic evolution, when matched with a Cauchy interior 
evolution, should perform well.

\section{3D Adaptive Mesh Refinement}
3D BH simulations are very demanding computationally.  In this 
section I outline the computational needs, and techniques designed to 
reduce them.  We will need to resolve waves with wavelengths of order 
$5M$ or less, where $M$ is the mass of the BH.  Although for 
Schwarzschild, the fundamental $\ell=2$ quasinormal mode wavelength is 
$16.8M$, higher modes, such as $\ell=4$ and above, have wavelengths of 
$8M$ and below.  The BH itself has a radius of $2M$.  More 
important, for very rapidly rotating Kerr BH's, which are 
expected to be formed in realistic astrophysical BH 
coalescence, the modes are shifted down to significantly shorter 
wavelengths\cite{Flanagan97a,Flanagan97b}.  As we need of order 20 
grid zones to resolve a single wavelength, we can conservatively 
estimate a required grid resolution of about $\Delta x = \Delta y = 
\Delta z \approx 0.2M$.  For simulations of time scales of order $t 
\propto 10^{2}-10^{3}M$, which will be required to follow coalescence, 
the outer boundary will probably be placed at a distance of roughly $R 
\propto 100M$ from the coalescence, requiring a Cartesian simulation 
domain of about $200M$ across.  This leads to about $10^{3}$ grid 
zones in each dimension, or about $10^{9}$ grid zones in total.  As 3D 
codes to solve the full Einstein equations have typically 100 
variables to be stored at each location, and simulations are performed 
in double precision arithmetic, this leads to a memory requirement of 
order 1000 Gbytes!  (In fairness to some groups that use spectral 
methods instead of finite differences (e.g., the Meudon group), I 
should point out highly accurate 3D simulations can now be achieved on 
problems that are well suited to such techniques, using much less 
memory!~\cite{Bonazzola98a}).

The largest supercomputers available to scientific research 
communities today have only about $\frac{1}{20}$ of this capacity, and 
machines with such capacity will not be available for some years.  
Furthermore, if one needs to double the resolution in each direction 
for a more refined simulation, the memory requirements increase by an 
order of magnitude.  Although such estimates will vary, depending on 
the ultimate effectiveness of inner or outer boundary treatments, 
gauge conditions, etc., they indicate that barring some unforeseen 
simplification, some form of adaptive mesh refinement (AMR) that 
places resolution only where it is required is not only desirable, but 
essential.  The basic idea of AMR is to use some set of criteria to 
evaluate the quality of the solution on the present time step.  If 
there are regions that require more resolution, then data are 
interpolated onto a finer grid in those regions; if less resolution is 
required, grid points are destroyed.  Then the evolution proceeds to 
the next time step on this hierarchy of grids, where the process is 
repeated.  These rough ideas have been refined and applied in many 
applications now in computational science.

There are several efforts ongoing in AMR for relativity.  Choptuik was 
the early pioneer in this area, developing a 1D AMR system to handle 
the resolution requirements needed to follow scalar field collapse to 
a BH\cite{Choptuik89}.  As an initially regular distribution of scalar 
field collapses, it will require more and more resolution as its 
density builds up.  The grid density required to resolve the initial 
distribution may not even see the final BH. Further, as pulses of 
radiation propagate back out from the origin, they, too may have to be 
resolved in regions where there was previously a coarse grid.  
Choptuik's AMR system, built on early work of Berger and 
Oliger\cite{Berger84}, was able to track dynamically features that 
develop, enabling him to discover and accurately measure BH critical 
phenomena that have now become so widely studied\cite{Choptuik93}.

Based on this success and others, and on the general considerations 
discussed above, full 3D AMR systems are under development to handle 
the much greater needs of solving the full set of 3D Einstein 
equations.  A large collaboration, begun by the Grand Challenge 
Alliance, has been developing a system for distributing computing on 
large parallel machines, called Distributed Adapted Grid Hierarchies, 
or DAGH. Among other things, DAGH provides a framework for parallel 
AMR, and is one of the major computational science accomplishments to 
come out of the Alliance.  Developed by Manish Parashar and Jim 
Browne, in collaboration with many subgroups within and without the 
Alliance, it is now being applied to many problems in science and 
engineering.  One can find information about DAGH online at 
http://www.cs.utexas.edu/users/dagh/.

\begin{figure}
\centerline{
\psfig{figure=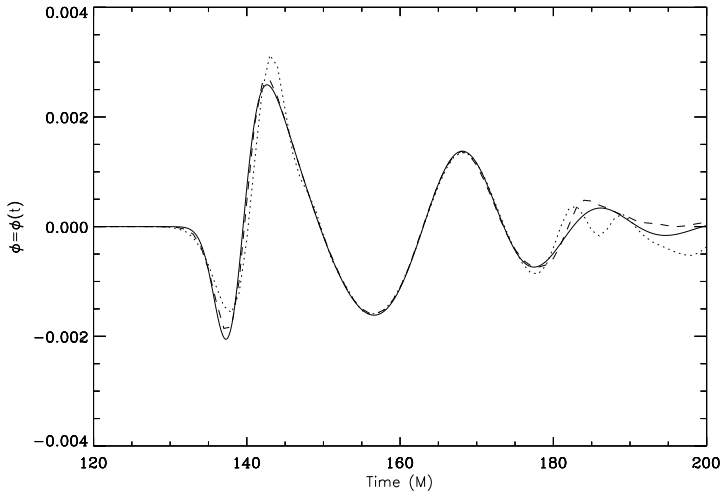,height=70mm}}
\caption{Tracking of outgoing waves using one and two levels of 
dynamic refinement.  The signal shown is seen by an equatorial 
observer located at 125M. The base grid resolution is 2M. Using one 
level of refinement captures and contains the first two outgoing modes 
(dotted line) compared to the the 1D result (solid line).  The quality 
of the signal improves even more when two levels of dynamic refinement 
are used (dashed line).}
\label{amazing}
\end{figure}

At least two other 3D software environments for AMR have been 
developed for relativity: one is called HLL, or Hierarchical Linked 
Lists, developed by Lee Wild and Bernard Schutz\cite{Wild98a}; 
another, called BAM, was the first AMR application in 3D relativity 
developed by Br{\"u}gmann~\cite{Bruegmann96}, and will be discussed 
later.  The HLL system has recently been applied to the test problem 
of the Zerilli equation describing perturbations of black 
holes\cite{Papadapoulos98a}.  As emphasized by Pullin in his GR15 
talk, this nearly 30 year old linear equation is still providing a 
powerful model for studying BH collisions, and it is also being used as a 
model problem for 3D AMR. In this work, the 1D Zerilli equation is 
recast as a 3D equation in cartesian coordinates, and evolved within 
the AMR system provided by HLL. Even though the 3D Zerilli equation is 
a single linear equation, it is quite demanding in terms of resolution 
requirements, and without AMR it is extremely difficult to resolve 
both the initial pulse of radiation, the blue shifting of waves as 
they approach the horizon, and the scattering of radiation, including 
the normal modes, far from the hole.  In Fig.~\ref{amazing} I show 
results obtained using this system.  The effect of the AMR is 
impressive, allowing one to capture the physics accurately even when 
the ``base grid'', which is the coarsest resolution level, is 
completely inadequate to resolve the physically interesting features.

\section{Outer Boundary Treatments}

Appropriate conditions for the outer boundary have yet to be derived 
for 3D. In 1D and 2D codes, the outer boundary is simply placed far 
enough away that the spacetime is nearly flat there, and static or 
flat boundary conditions can usually be specified for the evolved 
functions.  However, due to the constraints placed on us by limited 
computer memory, this is not currently possible in 3D. AMR will be of 
great use in this regard, but will not substitute for proper physical 
treatment.  Most results to date have been computed with the evolved 
functions kept static at the outer boundary, even if the boundaries 
are too close for comfort in 3D!

There are several other approaches under development that promise to 
improve this situation greatly that I will not have room to explore in 
detail here, but should be mentioned.  Generally, one has in mind 
using Cauchy evolution in the strong field, interior region where the 
BH's are colliding.  This outer part of this region will be 
matched to some exterior treatment designed to handle what is 
primarily expected to be outgoing radiation.  

Two major approaches have been developed by the Grand Challenge 
Alliance and other groups.  First, by using perturbation theory, as 
described later in this paper, it is possible to identify quantities 
in the numerically evolved metric functions that obey the 
Regge-Wheeler and Zerilli wave equations.  These can be used to 
provide boundary conditions on the metric and extrinsic curvature 
functions in an actual evolution, as described in a recent paper from 
the Grand Challenge Alliance~\cite{Abrahams97a}.  This is an excellent 
step forward in outer boundary treatments that should work to minimize 
reflections of the outgoing wave signals from the outer boundary.  In 
tests with weak waves, a full 3D Cauchy evolution code has been 
successfully matched to the perturbative treatment at the boundary, 
permitting waves to escape from the interior region with very little 
reflection.  Alternatively, ``Cauchy-Characteristic matching'' 
attempts to match spacelike slices in the Cauchy region to null slices 
at some finite radius, and the null slices can be carried out to 
$\cal{I}$.  As described above, the full 3D characteristic evolution 
codes have progressed dramatically in recent years, and although the 
full 3D matching remains to be completed, tests of the scheme in 
specialized settings show promise\cite{Bishop98a}.  One can also use 
the hyperbolic formulations of the Einstein equations to find 
eigenfields, for which outgoing conditions can in principle be 
applied\cite{Bona94b}.  In 3D this technique is still under 
development, but it exploits mathematical properties of the equations, 
and 1D tests work well, it shows promise for future work.  Finally, 
another hyperbolic approach uses conformal rescaling to move the 
boundary to 
infinity~\cite{Friedrich81a,Friedrich81b,Friedrich96,Huebner96}.  
These methods have different strengths and weaknesses, but all promise 
to improve boundary treatments significantly, helping to enable longer 
evolutions than are presently possible.

\section{3D Dynamic BH Simulations}
I now turn to what has actually been achieved over the last few years 
in actual 3D BH evolutions in a Cauchy evolution setting, which is 
expected to be the main line of attack for the general binary BH 
merger problem.  Although I have discussed many techniques above that 
are thought to be needed for the general problem, such as AMR, AHBC, 
advanced boundary treatments, and so on, in this section I discuss 
what has already been possible without such advanced algorithms.

In what follows, I discuss a series of simulations carried out in 3D 
cartesian coordinates with a fixed, 3D mesh (implying that resolution 
is very limited, even on the world largest supercomputers), with 
standard singularity avoiding slicings instead of AHBC (implying that 
slices will become pathologically warped, causing the codes to crash), 
and with fixed outer boundaries (implying that waves that reach the 
boundary will be reflected back into the domain of interest).  In 
spite of all of these caveats, we will see that already one can 
achieve quite remarkable results in 3D, which can be verified through 
a series of testbed and convergence calculations.  As advanced 
algorithms are developed, they will be tested on simulations such as 
these, and should extend our capabilities with each step forward.

\subsection{Distorted BH's:  3D Spectroscopy}
I begin with a simulation of a distorted single BH in a 3D 
code, with an initial data set of the ``Brill wave plus BH'' 
type discussed above.  One can consider this as a prototype of a black 
hole just formed during the collision process of two merging black 
holes.  The goal here is to see if one can evolve it properly in a 
full 3D code, track the waves emitted as it settles down, and extract 
them from the metric functions actually being evolved.

\begin{figure}
\centerline{\psfig{figure=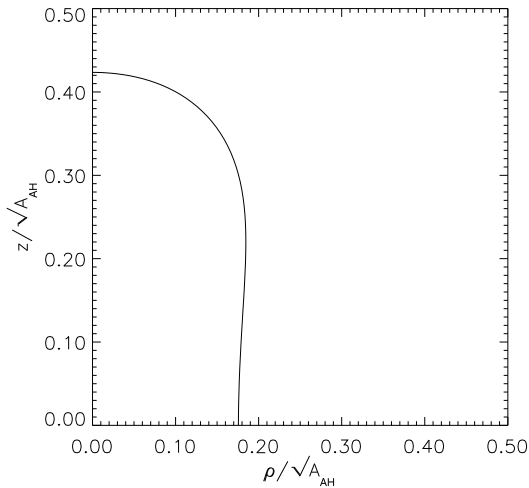,height=70mm}} 
\caption{We show the embedding diagram for the apparent horizon of an 
axisymmetric, highly distorted BH initial dataset.  Embedding 
coordinates are normalized by the square root of the area.  Such a BH 
is similar in shape to BH's formed during the head-on collision of two 
black holes, and is a useful test case for 3D numerical evolution.}
\label{embedding}
\end{figure}

As an example of the type of initial data under consideration, I first 
show in Fig.~\ref{embedding} an embedding diagram of the apparent 
horizon of such a hole.  In this case, I show an axisymmetric hole, 
because the horizon embeddings are easy to compute, but below I will 
consider evolutions for both axisymmetric and full 3D BH 
initial data.

The 3D code, developed originally by the NCSA/WashU/Potsdam 
collaboration, and developed further for these simulations by Karen 
Camarda, is written without making use of any symmetry assumptions.  
The code is a general 3D ADM code (the so-called ``G'' code), allowing 
very general slicings and shift conditions, but the particular 
simulations shown here use zero shift and a particular singularity 
avoiding slicing described in Ref.~\cite{Camarda97b,Camarda97c}.  The 
initial data I discuss here have both equatorial plane symmetry and 
quadrant symmetry (i.e., although fully 3D, any intrinsic 
$\phi-$dependence is repeated in each quadrant).  Hence we can save on 
the memory and computation required by evolving only one octant of the 
system.  As discussed above, without some form of memory savings, 
highly resolved, 3D simulations with outer boundaries sufficiently far 
away are simply not possible on even the largest available computers 
in 1997.  As shown in \cite{Anninos94c}, this trick has no effect on 
the simulations except to reduce the computational requirements by a 
factor of eight.  Even with such computational savings, these are 
extravagant calculations!  The results presented in this paper were 
computed on a 3D Cartesian grid of $300^{3}$ numerical grid zones, 
take about 12 Gbytes of memory, and require about a day on a 128 
processor, SGI/Cray Origin 2000 parallel supercomputer.

The questions we want to answer with these simulations are: 
(a) Can we evolve highly distorted BH's, like those formed in 
a collision, in a general 3D simulation code?; (b) Can we extract 
radiation, even when the waves are very weak, with energy $E < 
10^{-3} M$?;  (c) Do we know if we get the right answer?    The 
answer to all three questions is an emphatic YES!.  By using a 
combination of 2D codes and perturbative testbeds, we will see that 
even very weak $\ell-$modes, including nonaxisymmetric 
$\ell=4$ modes, can be very accurately obtained in a full 3D cartesian
simulation.  For this reason, I like to refer to this as BH 
spectroscopy!  Many energy levels of the BH excitations 
(quasinormal modes) can be followed and studied in full 3D.

There are many ways to evolve such a distorted BH 
system, and I will discuss and compare three of them here:  (a)  
perturbative evolution, (b) axisymmetric evolution in the case where 
there is no $\phi$ dependence, and (c) full 3D evolution as above.

\subsubsection{Comparison with results from mature 2D codes.}
Over the last decade, very mature 2D codes have been developed and 
well tested.  These codes have been applied to distorted 
Schwarzschild~\cite{Abrahams92a}, Misner colliding black 
holes~\cite{Anninos93b,Anninos94a}, and distorted rotating black 
holes~\cite{Brandt94c}.  They provide an excellent testing ground for 
full 3D evolutions, as one can transform the initial data sets into 
Cartesian coordinates, and evolve them as full 3D data sets, even 
though the underlying initial data are axisymmetric.  As the 2D and 3D 
codes use completely different coordinate systems, gauges, slicings, 
etc., even the metric functions that are evolved will be very 
different: only the physics should be the same in both codes.

One particular measure of the physics, which is most appropriate for 
gravitational wave astronomy, is a waveform seen by a distant 
observer.  This can be computed using an extraction technique 
developed originally by Abrahams\cite{Abrahams88,Abrahams90}.  This 
technique is based on a gauge-invariant perturbation theory developed 
by Moncrief~\cite{Moncrief74}, and in the present 3D application is 
detailed in Refs.\cite{Camarda97a,Camarda97c,Allen97a}.  Essentially, 
the Zerilli function $\psi$, which obeys the Zerilli wave equation 
discussed above, is computed as a function of time at various radii 
away from the distorted BH.

\begin{figure}
\centerline{
\psfig{figure=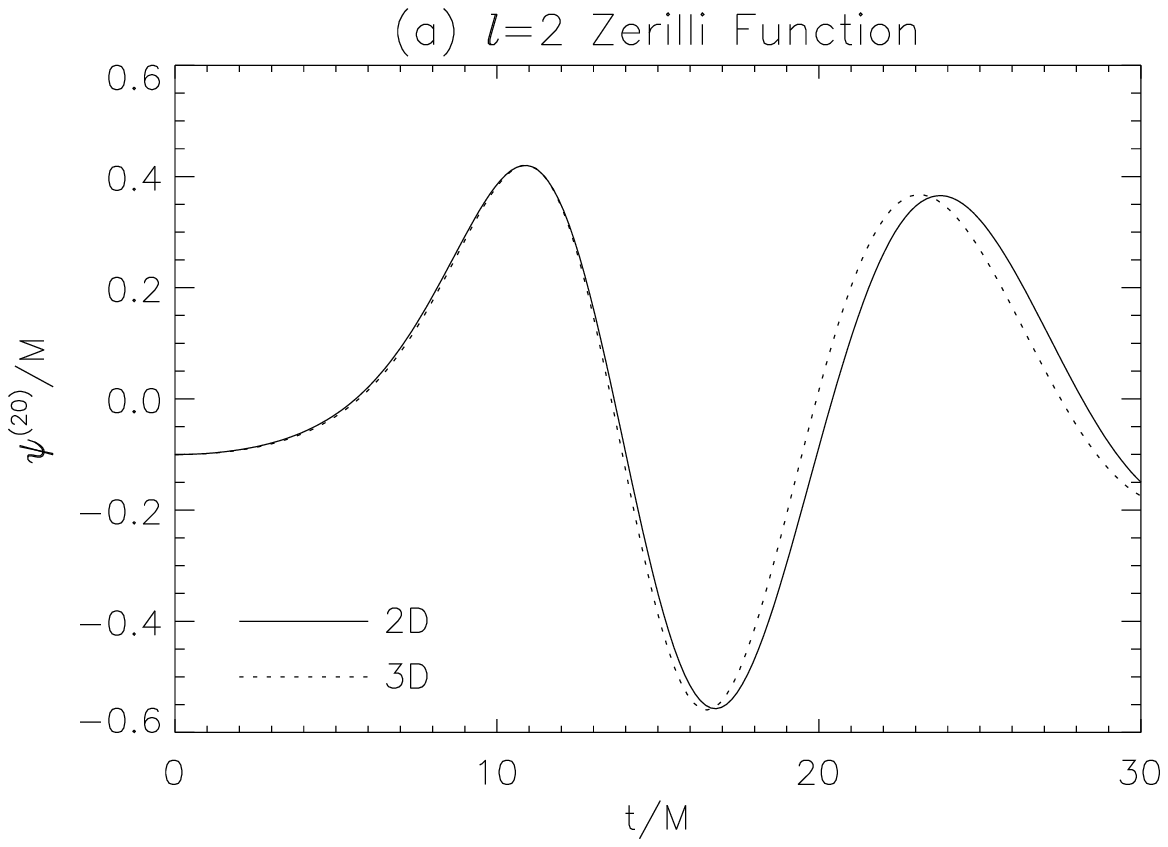,height=70mm}}
\centerline{\psfig{figure=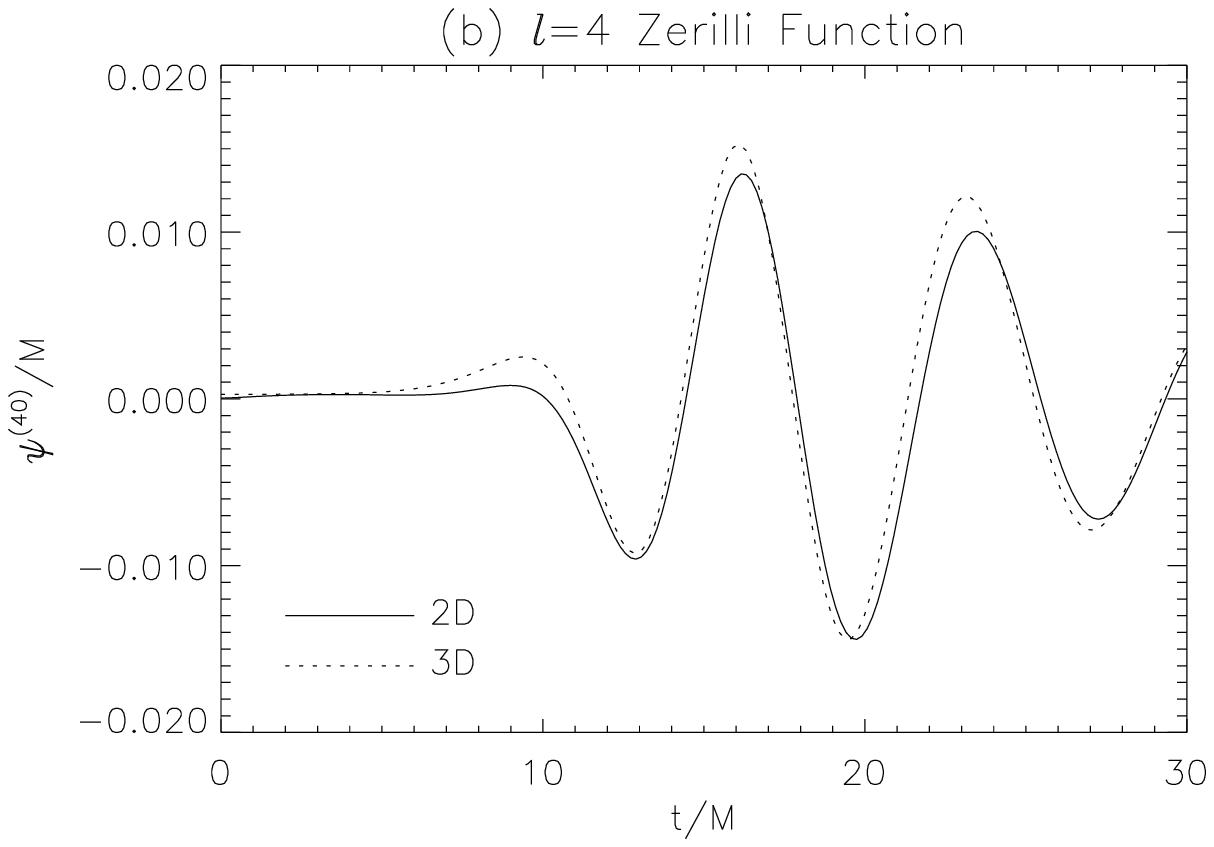,height=70mm}}
\caption{We show the (a) $\ell=2$ and (b) $\ell=4$ Zerilli functions
vs. time, extracted during 2D and 3D evolutions of the data set
$(a,b,w,n,c)=(0.5,0,1,2,0)$. The functions were extracted at a radius of
$8.7M$. The 2D data were obtained with $202\times 54$ grid points,
giving a resolution of $\Delta\eta=\Delta\theta=0.03$. The 3D data
were obtained using $300^3$ grid points and a resolution of $\Delta
x=0.0816M$.}
\label{fig:zer2dcomp}
\end{figure}

As an example of such simulations, we study the evolution of the 
distorted single BH initial data set, similar to the one whose horizon 
embedding is shown above ($(a,b,w,n,c) = (0.5,0,1,2,0)$ in the 
language of Ref.~\cite{Camarda97c}).  In Fig.~\ref{fig:zer2dcomp}a we 
show the result of the 3D evolution, focusing on the $\ell=2$ Zerilli 
function extracted at a radius $r=8.7M$ as a function of time.  
Superimposed on this plot is the same function computed during the 
evolution of the same initial data set with a 2D code, based on the 
one described in detail in \cite{Abrahams92a,Bernstein93b}.  The 
agreement of the two plots is quite remarkable.  It is important to 
emphasize that the two results were computed with different slicings, 
different coordinate systems, and {\em different spatial gauges}.  Yet 
the physical results obtained by these two different numerical codes, 
as measured by the waveforms, are remarkably similar (as one would 
hope).  A full evolution with the 2D code to $t=100M$, by which time 
the hole has settled down to Schwarzschild, shows that the energy 
emitted in this mode at that time is about $4\times 10^{-3}M$.  This 
result shows that now it is possible in full 3D numerical relativity, 
in cartesian coordinates, to study the evolution and waveforms emitted 
from highly distorted BH's, even when the final waves leaving the 
system carry a small amount of energy.

In Fig.~\ref{fig:zer2dcomp}b we show the $\ell=4$ Zerilli function 
extracted at the same radius, computed during evolutions with 2D and 
3D codes.  This waveform is more difficult to extract, because it has 
a higher frequency in both its angular and radial dependence, and it 
has a much lower amplitude: the energy emitted in this mode is three 
orders of magnitude smaller than the energy emitted in the $\ell=2$ 
mode, {\em i.e.}, $10^{-6}M$, yet it can still be accurately evolved 
and extracted.  This is quite a remarkable result, and bodes well for 
the ability of numerical relativity codes ultimately to compute 
accurate waveforms, which are buried deeply in the metric functions 
actually evolved, that will be of great use in interpreting data 
collected by gravitational wave detectors.  (However, as I point out 
below, there is a quite a long way to go before the general 3D 
coalescence can be studied!)

\subsubsection{Comparison against full 3D perturbative evolution}
After passing tests of 3D evolution of axisymmetric distorted black 
hole initial data, we now turn to full 3D distorted BH data sets, for 
which there are no axisymmetric treatments available for comparison.  
However, if distortions are fairly small, one expects that the initial 
data can be evolved by perturbation theory.  As Pullin describes in 
detail in this volume, this approach has been remarkably successful in 
handling a variety of BH systems.  The approach is similar to that 
used above to extract the waveforms, except that in this case the 
Zerilli function is computed throughout the spatial domain in the back 
hole initial data.  This provides Cauchy data for the Zerilli 
evolution equation, which can then be used to evolve all $\ell-$modes 
forward in time.  The results can then be compared with the full 
nonlinear evolution, which is analyzed using the gauge-invariant 
waveform extraction procedure described above.  If all is well, and 
the evolutions are truly in the perturbative regime, the results 
should agree.

\begin{figure}
\centerline{
\psfig{figure=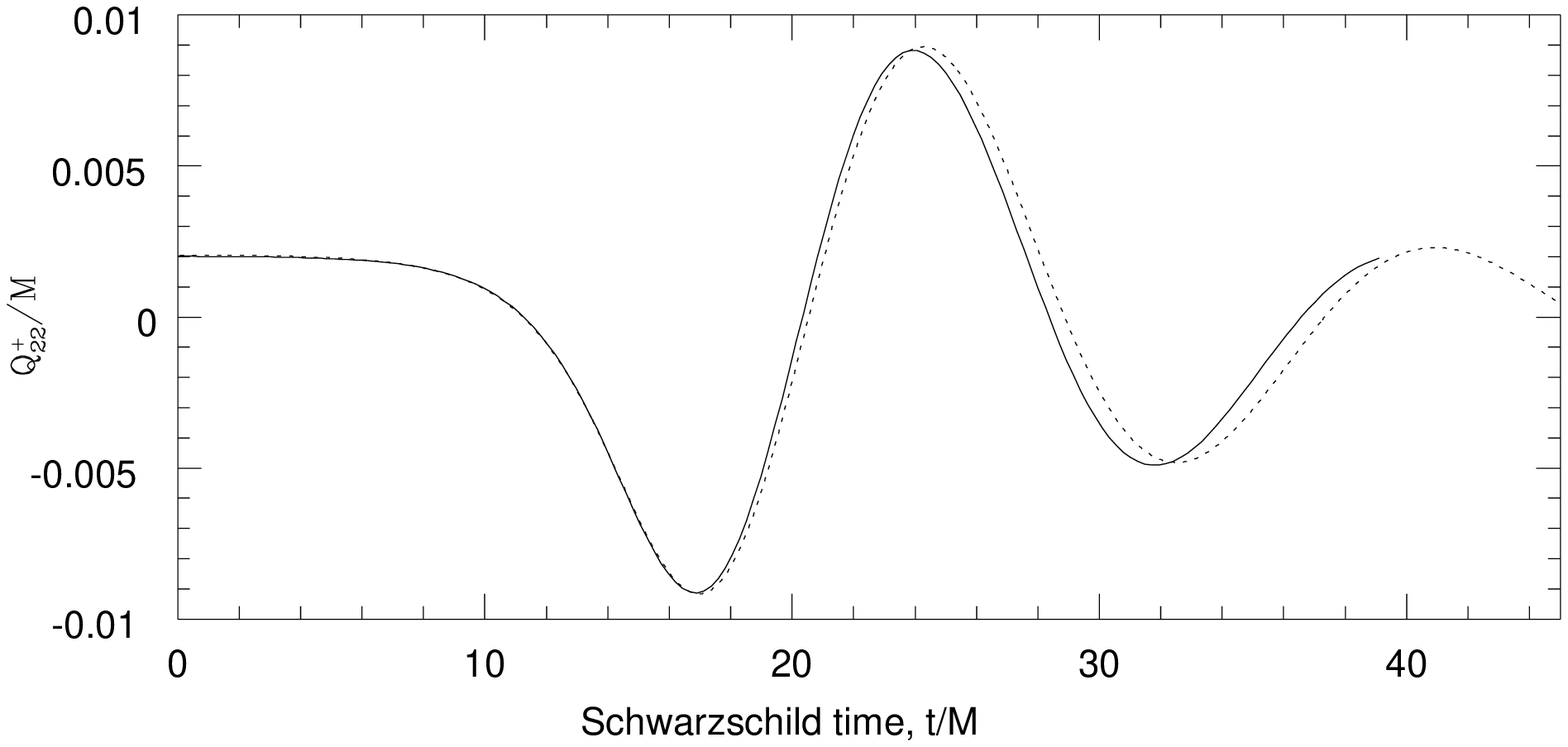,height=70mm}}
\centerline{\psfig{figure=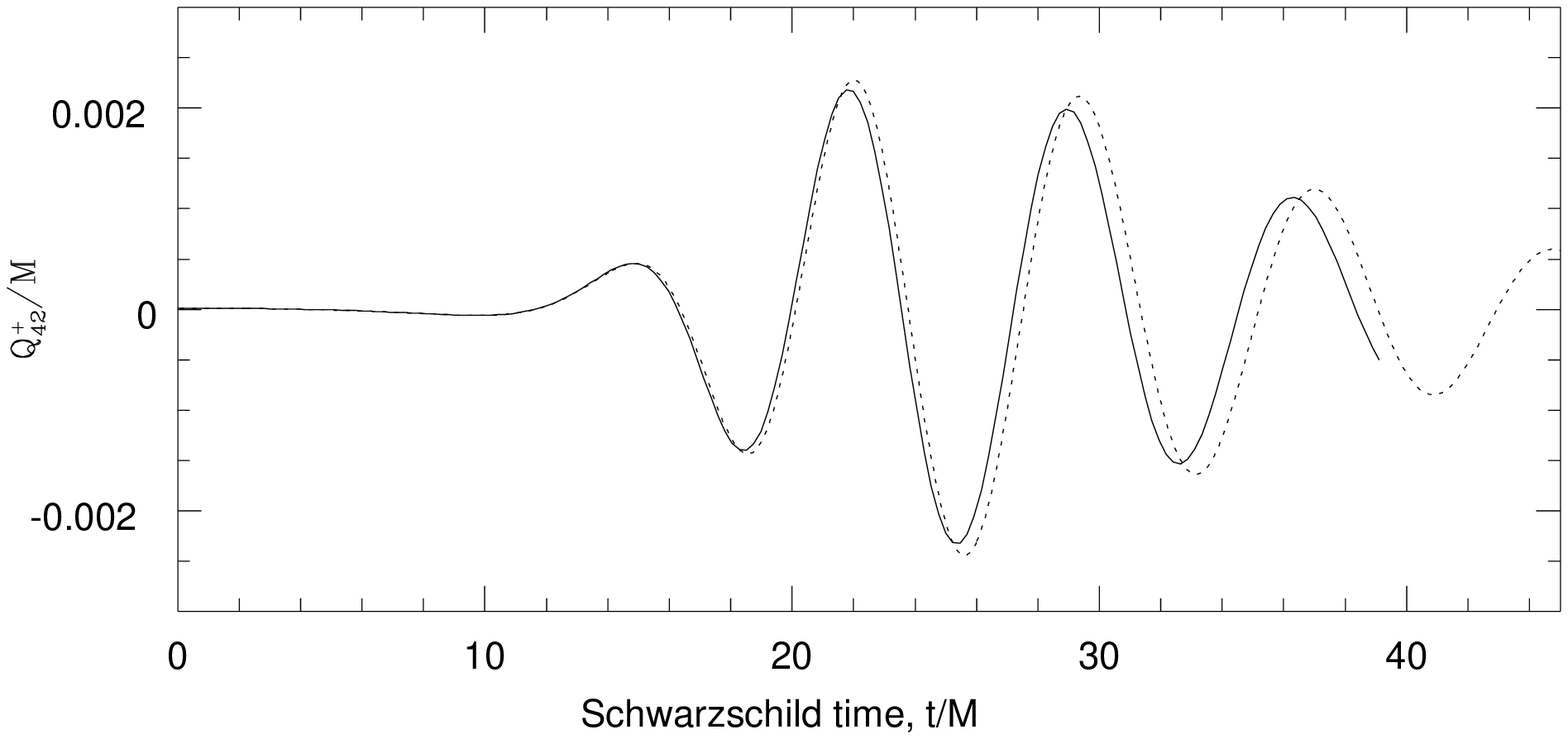,height=70mm}}
\caption{We show waveforms for the (a) $\ell=m=2$  and (b) 
$\ell=4,m=2$ nonaxisymmetric modes
extracted from the full nonlinear simulation of a 3D distorted BH. 
Solid lines show the nonlinear evolutions, and the dotted lines show 
the perturbative results.}
\label{fig:3dmode}
\end{figure}

In Fig.~\ref{fig:3dmode} I show the results of one such comparison.  A 
3D BH is evolved with the full 3D nonlinear code described 
above.  The waveform is extracted from the simulation, and compared to 
the results of the perturbative evolution.  The mode shown in 
Fig.~\ref{fig:3dmode}a is the nonaxisymmetric $\ell=m=2$ mode, already 
described above as one of the most relevant for gravitational wave 
astronomy.  The waveform in Fig.~\ref{fig:3dmode}b is the higher order 
$\ell=4,m=2$ mode, which carries much lower energy.  These results 
have been reported in much more detail in 
\cite{Camarda97a,Camarda97b,Allen97a,Allen98a}.

To summarize these results: In recent years great progress has been 
made in full 3D numerical relativity applications to BH evolutions.  
We can now evolve 3D distorted BH's, with standard slicing techniques, 
long enough to track the development of the radiation patterns emitted 
during the ringdown of the BH. This is the first time that true 3D 
BH's have been evolved in full numerical relativity, and the 
perturbative results confirm that even the minute details of the 
spectrum of gravitational radiation emitted, carrying energy of order 
$10^{-6}M$, are accurate.  Although there are still many long steps to 
the general coalescence problem, for this class of test problems, I think 
it is fair to say that 3D numerical relativity has progressed from 
blunt instrument to fine art: 3D BH spectroscopy is now possible!

\subsection{First 3D Collision of 2 BH's}

Now I move on to the problem of two colliding BH's, which is the long 
term goal.  This is a much harder problem that will ultimately require 
the advanced techniques under development, such as AHBC, AMR, advanced 
BC's, etc, but as always there are simpler stepping stones to the 
general merger system.  We take the Misner data as our prototype BH 
collision, and see what is possible in 3D. As discussed above, the 
Misner two BH data has played a central role in numerical relativity 
for more than three decades.  Through extensive axisymmetric 
simulations~\cite{Anninos93b,Anninos94a,Anninos98a}, perturbation 
theory (Pullin's lecture), and horizon studies\cite{Anninos94f}, this 
is a true two BH system that is understood in great detail.
	
\begin{figure}
\centerline{\psfig{figure=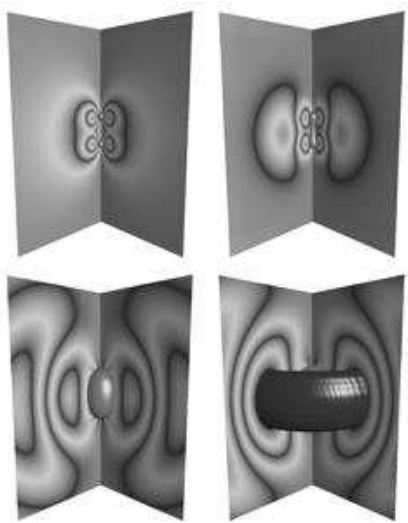,height=70mm}} 
 \caption{3D Evolution of the head-on collision of two black holes.  
 The radiation field $\Psi_4$ is shown as a grayscale map.  The event 
 horizon is shown as a solid object in the center.  Features compared 
 to the results obtained in 2D evolutions show good agreement.}
\label{3dpsi4}
\end{figure}

We have also computed the head-on collision of two equal mass black 
holes in the 3D code.  Preliminary results agree very well with 2D, 
although we cannot yet evolve the 3D system as far into the future.  
In Fig.~\ref{3dpsi4} I show the evolution of the radiation field 
$\Psi_4$ as a grayscale map, and the coordinate position of the event 
horizon, traced out using the techniques described above.  Notice the 
``banana'' shaped quadrupole lobes of radiation propagating out from 
the colliding holes, just as in the 2D calculations.  Quantitative 
studies of the coalescence time of the horizons also show excellent 
agreement with the 2D studies\cite{Anninos96c}.

This work is already more than two years old, but shows what is 
possible at present even without advanced techniques such as AHBC and 
AMR, and that for highly dynamic colliding BH spacetimes, 3D 
calculations are capable of producing waveforms and horizon dynamics.  
These calculations are now being redone with new codes (see below), 
and bigger computers, and should yield more accurate and detailed 
results.  Further, 3D calculations such as these will provide 
important testbeds for the more advanced techniques as they are 
developed.

This is exciting progress, but there is still a long way to go!  Up 
to this point, important features, such as orbital angular momentum, 
have not been considered.  We turn to the general binary merger 
case next. 

\subsection{First true 3D BH Collision Simulation}
The first attempt to test out the general 3D binary BH data in an 
evolution code was recently made by Br{\"u}gmann~\cite{Bruegmann97}.  
Using an ADM 3D code (BAM, independent of the one used in the above 
simulations), he recently evolved a true 3D binary BH dataset, with 
spin and angular momentum, going beyond single distorted 3D BH's and 
simplified axisymmetric BH collisions.  The datasets he evolved belong 
to the new family of ``Black Hole Punctures''~\cite{Brandt97b}, the 
generalization of multiple Schwarzschild holes with singularities, as 
described above.

As in the above simulations, he used a ``traditional'' evolution 
approach: a 3D Cartesian grid, no shift, maximal slicing to avoid 
singularities, no AHBC, and fixed outer boundaries.  As discussed 
above, such simulations are extremely demanding computationally.  The 
results of the preceding section were achieved by making use of 
certain symmetries to reduce the computational domain required, but 
with these general data sets, no such reduction is possible.  The 
entire domain must be evolved.  In this case, one must resort to some 
form of adaptive computation in order to reasonably resolve the BH's 
and place the boundary reasonably far away.

Rather than employing a fully adaptive grid, which requires still some 
development, he employed a series of nested grids, each interior grid 
having higher resolution than the one that contains it.  This way one 
can achieve high resolution in the central region where BH's 
are merging, while placing the boundaries far away, in regions where 
one can afford to have rather coarse resolution.  Without such 
techniques, these calculations would be impossible.  Another innovative 
feature of this work is the coupling of maximal slicing, an elliptic 
equation, to the evolution equations, in the presence of nested grids.  
This a very difficult computational problem, and is perhaps the first 
successful implementation in 3D relativity.

\begin{figure}
\centerline{\psfig{figure=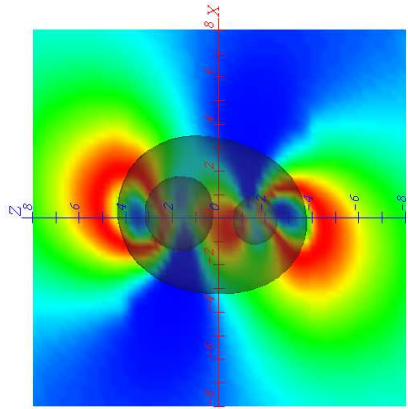,height=70mm}} 
\caption{A snapshot of true 3D binary BH evolution, showing the 
merging of two apparent horizons, shown inside the final horizon, the 
transparent surface engulfing them.  The grayscale map shows the 
metric function $g_{zz}$.}
\label{bruegmann}
\end{figure}

The results show the strength of this technique:  although the 
simulations could not be followed far into the future, it was possible 
to determine the location of the initial 3D apparent horizons, and to 
track the development of a global apparent horizon, indicating that 
the individual holes had merged, at a later time.  A snapshot of this 
simulation in shown in Fig.~\ref{bruegmann}, where one can see the 
two individual holes embedded in a larger horizon that developed 
towards the end of the simulation.

While very preliminary, this calculation gives a glimpse of what will 
be possible in the future.  It is reminiscent of the early 2D 
simulations of Smarr and Eppley~\cite{Smarr79}, when crude features of 
the Misner BH spacetime could be seen, but refined details, such as 
clean waveforms, would require still more development of numerical 
relativity techniques.  With each advance in algorithm technology, 
more sophisticated problems are being attacked, leading towards 
realistic astrophysical BH merger simulations.

\section{Putting the Pieces Together:  Codes for 3D Relativity}

As one can see, the solution to a single problem in numerical 
relativity requires a huge range of computational and mathematical 
techniques.  It is truly a large scale effort, involving experts in 
computer and computational science, mathematical relativity, 
astrophysics, and so on.  For these reasons, aided by collaborations 
such as the Grand Challenge Alliance, there has been a great focusing 
of effort over the last years.

A natural byproduct of this focusing has been the development of 
codes that are used and extended by large groups.  A code must have a 
large arsenal of modules at its disposal: different initial data sets, 
gauge conditions, horizon finders, slicing conditions, waveform 
extraction, elliptic equation solvers, AMR systems, boundary modules, 
different evolution modules, etc.  Furthermore, these codes must run 
efficiently on the most advanced supercomputers available.  Clearly, 
the development of such a sophisticated code is beyond any single 
person or group.  In fact, it is beyond the capability of a single 
community!  Different research communities, from computer science, 
physics, and astrophysics, must work together to develop such a code.

As an example of such a project, I describe briefly the ``Cactus'' 
code, developed by a large international collaboration\cite{Masso98a}.  
This code is an outgrowth of the last 5 years of 3D numerical 
relativity development primarily at NCSA/Potsdam/WashU, and builds 
heavily on the experience gained in developing the so-called ``G'' and 
``H'' codes~\cite{Anninos94c,Anninos94d,Masso98a}.  The core of Cactus 
was written from the ground up during 1997 by Paul Walker and Joan 
Mass\'o, and then heavily developed by the entire groups at Potsdam, 
WashU and NCSA. Presently, it is being developed collaboratively by 
these groups in collaboration with groups at Palma, Valencia, PRL in 
India, and computational science groups at U. of Illinois, and Argonne 
National Lab.

The code has a very modular structure, allowing different physics, 
analysis, and computational science modules to be plugged in.  In 
fact, versions of essentially all the modules listed above are already 
developed for the code.  For example, several formulations of 
Einstein's equations, including the ADM formalism and the Bona-Mass\'o 
hyperbolic formulation, can be chosen as input parameters, as can 
different gauge conditions, horizon finders, hydrodynamics evolvers, 
etc.  It is being tested on BH spacetimes, such as those described 
above, as well as on pure wave spacetimes, self-gravitating scalar 
fields and hydrodynamics.  It has also been designed to connect to 
DAGH ultimately for parallel AMR.

The code has also been heavily optimized to take advantage of the most 
powerful parallel supercomputers.  With help of experts at Cray and 
SGI, the code has recently achieved 100Gflops (100 billion floating 
point operations per second) on a 768 node Cray T3E, making it one the 
fastest general purpose production codes available in any area of 
scientific computing.

This code was also designed as a community code.  After first 
developing and testing it within our rather large community of 
collaborators, it will be made available with full documentation via 
a public ftp server maintained at AEI.  By having an entire research 
community using and contributing to such a code, we hope to 
accelerate the maturation of numerical relativity.  Information about 
the code is available online, and can be accessed at 
http://cactus.aei-potsdam.mpg.de.
									
\section{Summary}

To conclude, it is clear that 3D numerical relativity has had many 
successes over the last years, but that it also requires further 
development of basic algorithms before it will be able to solve fully 
such complex problems as the general merger of two spiraling black 
holes.  We have extensive families of BH initial data ready 
for evolution, and even with presently limited computational 
techniques it has been shown that highly accurate nonaxisymmetric 
waveforms can be obtained from simulations of fully 3D distorted black 
holes (black hole spectroscopy!)  and head-on collisions of black 
holes, and that one can already crudely study the merger of general 
binary BH's for limited times.  Further, characteristic 
evolution in 3D has made truly dramatic progress in the last year.

Extending our capabilities of highly accurate waveforms to true 3D 
BH mergers, with orbital angular momentum, will require the 
further development of advanced computational and algorithm 
techniques, including apparent horizon boundary conditions, adaptive 
mesh refinement, improved outer boundary conditions, perhaps through 
Cauchy-characteristic or perturbative matching, and a better 
understanding of gauge conditions (Gauge conditions are a 
major research area that I have not discussed, but one which will 
require a great deal of attention).  This is a tall order, but I have 
shown that in almost each area, dramatic progress has been made in the 
last few years.  AHBC has successfully employed general gauge 
conditions in one case to evolve a dynamic but nonmoving BH, 
and has also been used successfully to allow a boosted Schwarzschild 
hole to move across a 3D grid.  Full 3D AMR techniques have been 
demonstrated for model problems such as the 3D Zerilli equation to 
capture accurately the physics that would otherwise be unattainable 
with a 3D uniform grid code.  Large scale simulation codes, such as 
Cactus and the Grand Challenge Alliance codes, are under development 
by large collaborations, with the goal of integrating all these 
pieces for a unified attack on this problem.

I have discussed the important role played by testbeds in this work, 
but want to stress the powerful impact that collaborations with our 
colleagues in perturbation theory has had.  Fortunately, Jorge Pullin 
has covered this in his contribution.  I believe this rebirth of 
perturbative approaches to understanding BH interactions will 
continue to play a central in both the verification of numerical 
relativity and in the physical understanding and interpretation of the 
results.

I have focussed on black hole evolutions, and have had to leave out 
discussion of a large number of other topics central to numerical 
relativity that really deserve to be covered.  For example, there has 
been much talk about hyperbolic systems in numerical work over the 
last few years, and I regret not having space to discuss that here.  
The field is still very much alive, and the hopes that hyperbolic 
formulations will allow a superior numerical treatment and a deeper 
understanding of the Einstein equations are undamped.  In fact, a 
major motivation for the Cactus code was to provide a single framework 
for developing and comparing hyperbolic formulations with standard ADM 
formulations on a variety of problems, and I expect much work on this 
subject to continue to be published in the coming years.

Another major topic that has received no mention is work on coalescing 
neutron stars, another important source of gravitational waves.  
Several large scale efforts are underway to attack this problem, 
including a long term Japanese effort~\cite{Shibata97} and a NASA 
funded Grand Challenge effort involving researchers at 6 institutions 
in the US and Germany (http://wugrav.wustl.edu/nsnsgc/nsnsgc.html).  
The Cactus code is also playing a central role in the latter 
collaboration.

I hope it is clear that although there is much work to be done, 3D 
numerical relativity is improving rapidly, and that many exciting results 
are possible already, even with still limited computers and techniques 
available.  But even in those areas under development, we have a 
roadmap to address the problems we are facing, and the prognosis for 
improvement is excellent!

\section{Acknowledgments}
I would like to thank the organizers for inviting me to give this 
overview of current work in numerical relativity.  The work reviewed 
here in which I have been personally involved has been the result of a 
wonderful collaboration between the members of my groups at Illinois 
and Potsdam, the group led by Wai-Mo Suen at Washington University, 
and various other groups around the world.  Some of the work reviewed 
here was supported by grants NSF PHY/ASC 93--18152 (ARPA supplemented) 
and NASA-NCCS5-153.  Thanks also go to Miguel Alcubierre, Bernd 
Br{\"u}gmann, Harry Dimmelmeier, Gerd Lanfermann, Tom Goodale, and 
Ryoji Takahashi for carefully reviewing the paper.


\begin{thebibliography}{10}

\bibitem{Flanagan97b}
\'{E}anna \'{E}.~Flanagan and S.~A. Hughes, Phys. Rev. D {\bf 57},  4566
  (1998), gr-qc/9710129.

\bibitem{Flanagan97a}
\'{E}anna \'{E}.~Flanagan and S.~A. Hughes, Phys. Rev. D {\bf 57},  4535
  (1998), gr-qc/9701039.

\bibitem{Anninos94f}
P. Anninos, D. Bernstein, S. Brandt, J. Libson, J. Mass\'o, E. Seidel, L.
  Smarr, W.-M. Suen, and P. Walker, Phys. Rev. Lett. {\bf 74},  630  (1995).

\bibitem{Matzner95a}
R. Matzner, E. Seidel, S. Shapiro, L. Smarr, W.-M. Suen, S. Teukolsky, and J.
  Winicour, Science {\bf 270},  941  (1995).

\bibitem{Shapiro95a}
S. Shapiro, S. Teukolsky, and J. Winicour, Phys. Rev. D {\bf 52},    (1995).

\bibitem{York79}
J. York,  in {\em Sources of Gravitational Radiation}, edited by L. Smarr
  (Cambridge University Press, Cambridge, England, 1979).

\bibitem{Seidel96a}
E. Seidel,  in {\em Relativity and Scientific Computing}, edited by F. Hehl
  (Springer-Verlag, Berlin, Germany, 1996).

\bibitem{Seidel96b}
E. Seidel and W.-M. Suen,  in {\em Relativistic Gravitation and Gravitational
  Radiation}, edited by J.-P. Lasota and J.-A. Marck (Cambridge University
  Press, Cambridge, England, 1997), pp.\ 335--360.

\bibitem{Seidel97a}
E. Seidel,  in {\em Gravitation and Cosmology}, edited by S. Dhurandhar and T.
  Padmanabhan (Kluwer Academic, Dordrecht, 1997), pp.\ 125--144.

\bibitem{Seidel97b}
E. Seidel and W.-M. Suen, Acta Helvetica {\bf 69},  454  (1996).

\bibitem{Seidel98a}
E. Seidel,  in {\em On the Black Hole Trail}, edited by B. Iyer and B. Bhawal
  (Kluwer, 1998).

\bibitem{Brill59}
D.~S. Brill, Ann. Phys. {\bf 7},  466  (1959).

\bibitem{Camarda97b}
K. Camarda and E. Seidel, Phys. Rev. D {\bf 57},  R3204  (1998), gr-qc/9709075.

\bibitem{Brandt97a}
S. Brandt, K. Camarda, and E. Seidel, in preparation for Phys. Rev. D.

\bibitem{Brandt97c}
S. Brandt, K. Camarda, and E. Seidel,  in {\em Proc. 8th M. Grossmann Meeting},
  edited by T. Piran (World Scientific, Singapore, 1998), in press.

\bibitem{Bowen80}
J. Bowen and J.~W. York, Phys. Rev. D {\bf 21},  2047  (1980).

\bibitem{Misner60}
C. Misner, Phys. Rev. {\bf 118},  1110  (1960).

\bibitem{Brill63}
D.~S. Brill and R.~W. Lindquist, Phys. Rev. {\bf 131},  471  (1963).

\bibitem{Hahn64}
S.~G. Hahn and R.~W. Lindquist, Ann. Phys. {\bf 29},  304  (1964).

\bibitem{Smarr77}
L. Smarr, Ann. N. Y. Acad. Sci. {\bf 302},  569  (1977).

\bibitem{Cook93}
G.~B. Cook, M.~W. Choptuik, M.~R. Dubal, S. Klasky, R.~A. Matzner, and S.~R.
  Olivera, Phys. Rev. D {\bf 47},  1471  (1993).

\bibitem{Cook94}
G.~B. Cook, Phys. Rev. D {\bf 50},  5025  (1994).

\bibitem{Brandt97b}
S. Brandt and B. Br\"ugmann, Phys. Rev. Lett. {\bf 78},  3606  (1997).

\bibitem{Smarr78b}
L. Smarr and J. York, Phys. Rev. D {\bf 17},  2529  (1978).

\bibitem{Shapiro86}
S.~L. Shapiro and S.~A. Teukolsky,  in {\em Dynamical Spacetimes and Numerical
  Relativity}, edited by J.~M. Centrella (Cambridge University Press,
  Cambridge, England, 1986), pp.\ 74--100.

\bibitem{Bernstein89}
D. Bernstein, D. Hobill, and L. Smarr,  in {\em Frontiers in Numerical
  Relativity}, edited by C. Evans, L. Finn, and D. Hobill (Cambridge University
  Press, Cambridge, England, 1989), pp.\ 57--73.

\bibitem{Unruh84}
J. Thornburg, Classical and Quantum Gravity {\bf 14},  1119  (1987), unruh is
  cited here by Thornburg as originating AHBC.

\bibitem{Libson94a}
J. Libson, J. Mass\'o, E. Seidel, W.-M. Suen, and P. Walker, Phys. Rev. D {\bf
  53},  4335  (1996).

\bibitem{Seidel92a}
E. Seidel and W.-M. Suen, Phys. Rev. Lett. {\bf 69},  1845  (1992).

\bibitem{Alcubierre94a}
M. Alcubierre and B. Schutz, J. Comp. Phys. {\bf 112},  44  (1994).

\bibitem{Anninos94e}
P. Anninos, G. Daues, J. Mass\'o, E. Seidel, and W.-M. Suen, Phys. Rev. D {\bf
  51},  5562  (1995).

\bibitem{Scheel94}
M.~A. Scheel, S.~L. Shapiro, and S.~A. Teukolsky, Phys. Rev. D {\bf 51},  4208
  (1995).

\bibitem{Marsa96}
R. Marsa and M. Choptuik, Phys Rev D {\bf 54},  4929  (1996).

\bibitem{Anninos94c}
P. Anninos, K. Camarda, J. Mass\'o, E. Seidel, W.-M. Suen, and J. Towns, Phys.
  Rev. D {\bf 52},  2059  (1995).

\bibitem{Bruegmann96}
B. Br\"ugmann, Phys. Rev. D {\bf 54},    (1996).

\bibitem{Daues96a}
G.~E. Daues, Ph.D. thesis, Washington University, St. Louis, Missouri, 1996.

\bibitem{Cook97a}
G.~B. Cook {\it et~al.},   (1997), gr-qc/9711078.

\bibitem{Gomez97a}
R. Gomez, L. Lehner, R. Marsa, and J. Winicour,   (1997), gr-qc/9710138.

\bibitem{Gomez98a}
R. Gomez {\it et~al.},   (1998), gr-qc/9801069.

\bibitem{Bonazzola98a}
S. Bonazzola, E. Gourgoulhon, and J.-A. Marck,   (1998), astro-ph/9803086.

\bibitem{Choptuik89}
M. Choptuik,  in {\em Frontiers in Numerical Relativity}, edited by C. Evans,
  L. Finn, and D. Hobill (Cambridge University Press, Cambridge, England,
  1989).

\bibitem{Berger84}
M. Berger and J. Oliger, Journal of Computational Physics {\bf 53},  484
  (1984).

\bibitem{Choptuik93}
M. Choptuik, Phys. Rev. Lett. {\bf 70},  9  (1993).

\bibitem{Wild98a}
L. Wild and B. Schutz,   (1998), in preparation.

\bibitem{Papadapoulos98a}
P. Papadapoulos, E. Seidel, and L. Wild, Physical Review D  (1998), in 
press, gr-qc/9802069.

\bibitem{Abrahams97a}
A.~M. Abrahams {\it et~al.}, Physical Review Letters {\bf 80},  1812  (1998),
  gr-qc/9709082.

\bibitem{Bishop98a}
  N. Bishop, R. Isaacson, R. Gomez, L. Lehner, B. Szilagyi, and J. 
  Winicour, in {\em On the Black Hole Trail}, edited by B. Iyer and B. 
  Bhawal (Kluwer, 1998), gr-qc/9801070.

\bibitem{Bona94b}
C. Bona, J. Mass\'o, E. Seidel, and J. Stela, Phys. Rev. Lett. {\bf 75},  600
  (1995).

\bibitem{Friedrich81a}
H. Friedrich, Proc. Roy. Soc. London {\bf A 375},  169  (1981).

\bibitem{Friedrich81b}
H. Friedrich, Proc. Roy. Soc. London {\bf A 378},  401  (1981).

\bibitem{Friedrich96}
H. Friedrich, Class. Quant. Grav. {\bf 13},  1451  (1996).

\bibitem{Huebner96}
P. H\"ubner, Phys. Rev. D {\bf 53},  701  (1996).

\bibitem{Camarda97c}
K. Camarda and E. Seidel, gr-qc/9805099. Submitted to Physical Review D.

\bibitem{Abrahams92a}
A. Abrahams, D. Bernstein, D. Hobill, E. Seidel, and L. Smarr, Phys. Rev. D
  {\bf 45},  3544  (1992).

\bibitem{Anninos93b}
P. Anninos, D. Hobill, E. Seidel, L. Smarr, and W.-M. Suen, Phys. Rev. Lett.
  {\bf 71},  2851  (1993).

\bibitem{Anninos94a}
P. Anninos, D. Hobill, E. Seidel, L. Smarr, and W.-M. Suen, Technical Report
  No.~24, National Center for Supercomputing Applications.

\bibitem{Brandt94c}
S. Brandt and E. Seidel, Phys. Rev. D {\bf 52},  870  (1995).

\bibitem{Abrahams88}
A. Abrahams, Ph.D. thesis, University of Illinois, Urbana, Illinois, 1988.

\bibitem{Abrahams90}
A. Abrahams and C. Evans, Phys. Rev. D {\bf 42},  2585  (1990).

\bibitem{Moncrief74}
V. Moncrief, Annals of Physics {\bf 88},  323  (1974).

\bibitem{Camarda97a}
K. Camarda, Ph.D. thesis, University of Illinois at Urbana-Champaign, Urbana,
  Illinois, 1998.

\bibitem{Allen97a}
G. Allen, K. Camarda, and E. Seidel,   (1998), in preparation for Phys. 
Rev. Lett.

\bibitem{Bernstein93b}
D. Bernstein, D. Hobill, E. Seidel, L. Smarr, and J. Towns, Phys. Rev. D {\bf
  50},  5000  (1994).

\bibitem{Allen98a}
G. Allen, K. Camarda, and E. Seidel,   (1998), in preparation for Phys. 
Rev. D.

\bibitem{Anninos98a}
P. Anninos and S.~R. Brandt, Phys. Rev. Lett  (1998), in preparation.

\bibitem{Anninos96c}
P. Anninos, J. Mass\'o, E. Seidel, and W.-M. Suen, Physics World {\bf 9},  43
  (1996).

\bibitem{Bruegmann97}
B. Br\"ugmann,   (1997), gr-qc/9708035.

\bibitem{Smarr79}
L. Smarr,  in {\em Sources of Gravitational Radiation}, edited by L. Smarr
  (Cambridge University Press, Cambridge, England, 1979), p.\ 245.

\bibitem{Masso98a}
C. Bona, J. Mass\'o, E. Seidel, and P. Walker, (1998),
  gr-qc/9804065. Submitted to Physical Review D.

\bibitem{Anninos94d}
P. Anninos, J. Mass\'o, E. Seidel, W.-M. Suen, and M. Tobias, Phys. Rev. D {\bf
  56},  842  (1997).
  
\bibitem{Shibata97}
M. Shibata, K. Oohara, and T. Nakamura, Prog. Theor. Phys. {\bf 98},    (1997),
  to appear.

\end{thebibliography}

\end{document}